\documentclass[fleqn,11pt]{article}
\usepackage[cp1251]{inputenc}
\usepackage{amsmath,amssymb}

\usepackage{amsfonts,amssymb,cite}
\usepackage{graphicx}

\def\noi{\noindent}

\newcommand{\Title}[1]{\noi {{\Large\bf #1}}\\[1ex]}

\newcommand{\Author}[2]{\noi{\bf #1}\\[2ex]\noi{\normalsize\it #2}\\}

\newcommand{\Abstract}[1]{\vskip 2mm \begin{center}
        \parbox{16.4cm}{\small\noi #1} \end{center}\medskip}
\newcommand{\foom}[1]{\protect\footnotemark[#1]}

\def\nqq{\hspace*{-2em}}

\def\Jl#1#2{#1 {\bf #2},\ }

\def\ApJ#1 {\Jl{Astroph. J.}{#1}}
\def\CQG#1 {\Jl{Class. Quantum Grav.}{#1}}
\def\DAN#1 {\Jl{Dokl. AN SSSR}{#1}}
\def\GC#1 {\Jl{Grav. Cosmol.}{#1}}
\def\GRG#1 {\Jl{Gen. Rel. Grav.}{#1}}
\def\JETF#1 {\Jl{Zh. Eksp. Teor. Fiz.}{#1}}
\def\JETP#1 {\Jl{Sov. Phys. JETP}{#1}}
\def\JHEP#1 {\Jl{JHEP}{#1}}
\def\JMP#1 {\Jl{J. Math. Phys.}{#1}}
\def\NPB#1 {\Jl{Nucl. Phys. B}{#1}}
\def\NP#1 {\Jl{Nucl. Phys.}{#1}}
\def\PLA#1 {\Jl{Phys. Lett. A}{#1}}
\def\PLB#1 {\Jl{Phys. Lett. B}{#1}}
\def\PRD#1 {\Jl{Phys. Rev. D}{#1}}
\def\PRL#1 {\Jl{Phys. Rev. Lett.}{#1}}

\def\lal{&&\nqq {}}

\def\beq{\begin{equation}}
\def\eeq{\end{equation}}
\def\bear{\begin{eqnarray}}
\def\bearr{\begin{eqnarray} \lal}
\def\ear{\end{eqnarray}}
\def\earn{\nonumber \end{eqnarray}}

\begin{document}
\thispagestyle{empty}
\twocolumn[

\Title{Stability of the cosmological system of degenerated scalar charged fermions and Higgs scalar fields. I. Mathematical model of linear plane perturbations. \foom 1}

\Author{Yu.G. Ignat'ev}
    {Institute of Physics, Kazan Federal University, Kremlyovskaya str., 18, Kazan, 420008, Russia}


\Abstract
 {A mathematical model of the evolution of plane perturbations in the cosmological statistical system of completely degenerated scalar-charged fermions with the Higgs scalar interaction is formulated. A complete closed system of differential equations describing the unperturbed state of a homogeneous and isotropic system and a system of self-consistent equations describing the evolution of small perturbations are constructed.
}
\bigskip

] 
\section*{Introduction}
 A complete mathematical model of the cosmological evolution of the classical vacuum scalar Higgs field is formulated and investigated, both by methods of qualitative analysis and numerical simulation in\cite{Ignat20}. A mathematical model of the cosmological evolution of a complex system consisting of a degenerated Fermi liquid of scalar charged fermions and coupled with them classical and phantom scalar fields is formulated and investigated in \cite{YuI_20}. There were discovered physically notable peculiarities of the cosmological model which are manifested, in particular, in presence of giant bursts of cosmological acceleration. With this work, we begin the study of the stability of such models with respect to plane perturbations in order to clarify the influence of the factor of scalar attraction of fermions on the formation of the spatial structure of the Universe. The main goal of this article is formulation of the correct mathematical model of the evolution of small plane perturbations in a homogeneous and isotropic cosmological scalar charged degenerate Fermi system. On the one hand, the interest towards the systems of degenerated fermions is explained by the simplicity of the corresponding mathematical model, which is convenient for research.  On the other hand, such systems are interesting from the point of view of modeling objects of dark cold matter.

\section{Mathematical Model of a Degenerated Fermi-system with Scalar Higgs Fermion Interaction}
Below we consider a simple model of the cosmological system based on a one-sorted degenerated statistical system of scalar charged fermions and a single scalar Higgs field (classical one or a phantom).

The strict macroscopic consequences of the kinetic theory are the transport equations, including the conservation law for a certain
vector current corresponding to the microscopic conservation law in the reactions of a certain fundamental charge $ {\ rm q} $ (if such a conservation law exists) --
\begin{equation}\label{1}
\nabla_i q n^i=0,
\end{equation}
as well as conservations laws of energy - momentum of a statistical system  \cite{Ignat14_2}:
\begin{equation}\label{2}
\nabla _{k} T_{p}^{ik} -\sigma\nabla ^{i} \Phi =0,
\mathrm{}\end{equation}
where $n^i$ is a numeric vector, $T^{ik}_p$ is a energy momentum tensor (EMT) of particles; $\sigma$ -- is a density of scalar charges.

Under the local thermodynamic equilibrium (LTE) conditions, the statistical system is locally isotropic and is described by the locally equilibrium distribution function of fermions:

\begin{equation}\label{8_0}
f^0_a={\displaystyle \frac{1}{\mathrm{e}^{(-\mu_a+(u,p))/\theta}+ 1}},
\end{equation}
where $\mu_a$ is a chemical potential, $\theta$ is a local temperature, $u^i$ is a unit timelike vector of the statistical system's dynamic velocity. The kinematic momentum of the particle $p^i$ lies on the effective mass surface:
\begin{equation}\label{8_1}
(p,p)=m^2_* \Rightarrow \tilde{p}^4=\sqrt{m^2_*+\tilde{p}^2},
\end{equation}
where $\tilde{p}^{(i)}$ are reference projections of a momentum's vector, $p^2$ -- is a square of physical momentum.

Thus, in case of LTE the macroscopic moments \eqref{8_0} take form of corresponding moments of an ideal fluid \cite{Ignat14_1}:
\begin{equation}\label{3}
n^i=n u^i,
\end{equation}
\begin{equation}\label{4}
T_p^{ik}=(\varepsilon_p+p_p) u^iu^k-p_pg^{ik},
\end{equation}
where it is
\begin{equation}\label{5}
(u,u)=1.
\end{equation}
From the normalization relation (\ref{5}) the following known identity law results:
\begin{equation}\label{6}
u^k_{~,i}u_k\equiv 0.
\end{equation}
Therefore the conservation laws (\ref{2}) could be reduced to the following form:
\begin{eqnarray}\label{2a}
(\varepsilon_p+p_p)u^i_{~,k}u^k=(g^{ik}-u^iu^k)(p_{p,k}+\sigma\Phi_{,k});&\\
\label{2b}
\nabla_k[(\varepsilon_p+p_p)u^k]=u^k(p_{p,k}+\sigma\Phi_{,k}),&
\end{eqnarray}
and the conservation law of a fundamental charge $\mathrm{G}$ \eqref{1} turns to:
\begin{equation}\label{2c}
\nabla_k (nu^k)=0.
\end{equation}

Macroscopic scalars for a degenerated single-component system of fermions ($\theta\to 0$) under LTE conditions have the form \cite{YuI_20}:
\begin{equation}\label{2_3}
n=\frac{1}{\pi^2}\varrho^3;
\end{equation}
\begin{equation}\label{2_3a}
{\displaystyle
\begin{array}{l}
\varepsilon_p = {\displaystyle\frac{m_*^4}{8\pi^2}}F_2(\psi);
\end{array}}
\end{equation}
\begin{equation}\label{2_3b}{\displaystyle
\begin{array}{l}
p_p ={\displaystyle\frac{m_*^4}{24\pi^2}}(F_2(\psi)-4F_1(\psi))
\end{array}}
\end{equation}
\begin{equation}\label{2_3c}{\displaystyle
\begin{array}{l}
\sigma={\displaystyle\frac{q\cdot m_*^3}{2\pi^2}}F_1(\psi),
\end{array}}
\end{equation}
where the dimensionless function $\psi$ is introduced

\begin{equation}\label{psi}
\psi=\frac{\varrho}{m_*},
\end{equation}
which equals to the ratio of the Fermi momentum $\varrho$ to effective mass of a fermion, and the following $F_1(\psi)$ and $F_2(\psi)$ functions are introduced to shorten the notation:
\begin{equation}\label{F_1}
F_1(\psi)=\psi\sqrt{1+\psi^2}-\ln(\psi+\sqrt{1+\psi^2});
\end{equation}
\begin{equation}
\label{F_2}
F_2(\psi)=\psi\sqrt{1+\psi^2}(1+2\psi^2)-\ln(\psi+\sqrt{1+\psi^2}).
\end{equation}
The functions $F_1(x)$ and $F_2(x)$ are odd:
\begin{equation}\label{F(-x)}
F_1(-x)=-F_1(x);\quad F_2(-x)=-F_2(x),
\end{equation}
and have the following asymptotics
\begin{eqnarray}\label{F,x->0}
\left.F_1(x)\right|_{x\to 0}\simeq \frac{2}{3}x^3; \left.F_2(x)\right|_{x\to 0}\simeq \frac{8}{3}x^3;\nonumber\\
\left.(F_2(x)-4F_1(x))\right|_{x\to 0}\simeq \frac{8}{5}x^5
\end{eqnarray}
-- charge's mass in the field is significantly greater than the Fermi momentum (nonrelativistic limit) and
\begin{eqnarray}\label{F,x->8}
\left.F_1(x)\right|_{x\to\infty}\simeq x|x|;\quad \left.F_2(x)\right|_{x\to\infty}\simeq 2x^3|x|.
\end{eqnarray}
-- Fermi momentum is significantly greater than charge's mass (ultrarelativistic limit). It is easy to verify the validity of the identities:
\begin{equation}\label{E_P_f}
\varepsilon_p+p_p\equiv \frac{m^4_*}{3\pi^2}\psi^3\sqrt{1+\psi^2}.
\end{equation}
\begin{equation}\label{E_3P_f}
\varepsilon_p-3p_p\equiv \frac{m^4_*}{2\pi^2}F_1(\psi).
\end{equation}
In addition, let us note the differential relations which will be useful further:
\begin{eqnarray}\label{dF}
\frac{dF_1(x)}{dx}=\frac{x^2}{\sqrt{1+x^2}}; \; \frac{dF_2(x)}{dx}=8x^2\sqrt{1+x^2};&\nonumber\\
\frac{d}{dx}x^3\sqrt{1+x^2}=3x^2\sqrt{1+x^2}+\frac{x^4}{\sqrt{1+x^2}}.&
\end{eqnarray}
Further, following \cite{YuI_20}, we assume:\footnote{Here we  consider this option due to the simplicity of calculations. We intend to return to the more complicated option $m_*=|q\Phi|$  in the nearest future.}
\begin{equation}\label{m_*(pm)}
m_*=q\Phi.
\end{equation}
Then, due to $\psi(-\Phi)=-\psi(\Phi)$  and the odd properties \eqref{F(-x)} of the formula \eqref{2_3a} -- \eqref{2_3c}  lead to the following transformation laws
\begin{eqnarray}\label{trans_eps}
\varepsilon_p(-\Phi)=-\varepsilon_p(-\Phi);\nonumber\\
p_p(-\Phi)=-p_p(\Phi);\, \sigma(-\Phi)=\sigma(\Phi).\
\end{eqnarray}

Further, %
\begin{equation}\label{T_{iks}}
T_s^{ik}=\frac{1}{8\pi}\biggl(e\Phi^{,i}\Phi^{,k}-
\frac{e}{2}g^{ik}\Phi_{,j}\Phi^{,j}+g^{ik} V(\Phi)\biggr)
\end{equation}
-- is a tensor of energy - momentum of a scalar field, where the indicator $e=+1$ for a classical scalar field and $e=-1$  for a phantom scalar field, $V(\Phi)$ -- is the potential energy of a scalar field:
\begin{equation}\label{Higgs}
V(\Phi)=-\frac{\alpha}{4} \left(\Phi^{2} -\frac{m^{2} }{\alpha} \right)^{2} ,
\end{equation}
$\alpha$ -- is a self-action constant, $m$ -- is a mass of scalar bosons.

The scalar field $\Phi$ is defined by the equation for a charged scalar field with a source:
\begin{eqnarray}\label{Eq_Phi_r}
e\square\Phi+V'_\Phi{\Phi}=-8\pi\sigma \Rightarrow\nonumber\\
e\square\Phi+\frac{m^2\Phi^2}{2}-\frac{\alpha\Phi^4}{4}=-8\pi\sigma,
\end{eqnarray}
where $\square\psi$ -- is a d'Alembert operator on the $g_{ik}$ metric. As a consequence of \eqref{2} and \eqref{Eq_Phi_r} the conservation law of the total EMT of the system  ``charged fermions + scalar field'' is identically fulfilled:
\begin{equation}\label{dTik=0}
\nabla_iT^{ik}=\nabla_i\bigl(T^{ik}_{p}+T^{ik}_s\bigr)\equiv0.
\end{equation}
To make the system of equations closed, it is required to add Einstein equations with a $\Lambda$ - term:
\begin{equation}\label{EqEinst}
G^i_k\equiv R^i_k-\frac{1}{2}R\delta^i_k=8\pi T^i_k+\Lambda\delta^i_k.
\end{equation}

\section{The Unperturbed Isotropic Distribution}
Let us consider the following spatially flat Friedmann metric as a backround metric
\begin{equation}\label{ds_0}
ds_0^2=a^2(\eta)(d\eta^2-dx^2-dy^2-dz^2),
\end{equation}
and as a background solution we consider a homogeneous isotropic distribution of matter, where all thermodynamic functions and scalar fields depend only on time. One can easily verify that $u^i=\delta^i_4$ turns equations (\ref{2a}) into identity laws while the system of equations (\ref{2b}) -- (\ref{2c}) %
is reduced to two equations:
\begin{equation}\label{7a1}
\dot{\varepsilon}_p+3\frac{\dot{a}}{a}(\varepsilon_p+p_p)=\sigma\dot{\Phi};
\end{equation}
\begin{equation}\label{7b1}
\dot{n} +3\frac{\dot{a}}{a}n=0,
\end{equation}
where $\dot{f}=df/d\eta$. It is shown in \cite{Ignat14_2} that the charge's  conservation law \eqref{7b1} is a direct consequence of the energy conservation law \eqref{7a1}, which is, in turn, reduced to the simple relation:
\begin{equation}\label{aP}
\eqref{7a1}\Rightarrow a\varrho=\mathrm{const}.
\end{equation}
Taking into account  \eqref{aP} and \eqref{m_*(pm)} let write the dimensionless function $\psi$ \eqref{psi} in explicit form:
\begin{equation}\label{psi(eta)}
\psi(\eta)=\frac{\beta}{a(\eta)\Phi(\eta)},\quad \biggl(\beta=\frac{\varrho}{q},\; a(0)\equiv=1\biggl).
\end{equation}

Further, the tensor of energy - momentum of a scalar field in an unperturbed state also takes the form of a tensor of energy - momentum of an ideal isotropic fluid:
\begin{equation} \label{MET_s}
T_{s}^{ik} =(\varepsilon_s +p_{s} )u^{i} u^{k} -p_s g^{ik} ,
\end{equation}
where it is:
\begin{eqnarray}\label{Es}
\varepsilon_s=\frac{1}{8\pi}\biggl(\frac{\mathrm{e}}{2}\frac{\dot\Phi^2}{a^2}+V(\Phi)\biggr);\\
\label{Ps}
p_{s}=\frac{1}{8\pi}\biggl(\frac{\mathrm{e}}{2}\frac{\dot\Phi^2}{a^2}-V(\Phi)\biggr),
\end{eqnarray}
so that:
\begin{equation}\label{e+p}
\varepsilon_s+p_{s}=\frac{\mathrm{e}}{8\pi}\frac{\dot{\Phi}^2}{a^2}.
\end{equation}
The equation of the unperturbed scalar field in the Friedmann metric takes the form:
\begin{equation}\label{Eq_S_eta}
\mathrm{e}\biggl(\ddot{\Phi}+2\frac{\dot{a}}{a}\dot{\Phi}\biggr)+a^2(m^2\Phi-\alpha\Phi^3)= -8\pi a^2\sigma_0(\eta),
\end{equation}
where it is
\begin{equation}\label{sigma_0}
\sigma_0(\eta)=\frac{q^4\Phi^3(\eta)}{2\pi^2}F_1(\psi_0(\eta)).
\end{equation}

Finally, the independent zero-order Einstein equations have the following form (see \cite{YuI_20}):
\begin{eqnarray}\label{Surf_Einst}
3H^2-\tilde{\Lambda}-\frac{q^4\Phi^4}{\pi}F_2(\psi)-\nonumber\\
\frac{e\dot{\Phi}^2}{2a^2}-\frac{m^2\Phi^2}{2}+\frac{\alpha\Phi^4}{4}=0;
\end{eqnarray}
\begin{equation}\label{11-44}
\dot{H}+\frac{e\dot{\Phi}^2}{2a}+\frac{4 a}{3\pi}m^4_*\psi^3\sqrt{1+\psi^2}=0,
\end{equation}
where $H(\eta)$ -- is the Hubble constant,
\begin{equation}\label{H}
H= \frac{\dot{a}}{a^2},
\end{equation}
\begin{equation}\label{tilde{Lambda}}
\tilde{\Lambda}=\Lambda-\frac{m^4}{4\alpha}.
\end{equation}

The equation \eqref{11-44}  is obtained as a difference of the Einstein equations for the components $^1_1 - ^4_4$. Thus, the complete system of equations describing the unperturbed cosmological system consists of the Fermi momentum conservation law \eqref{aP}, the field equation \eqref{Eq_S_eta} and the Einstein equation \eqref{11-44} together with the definition of the Hubble constant \eqref{11-44}. In this case, the equation  \eqref{Surf_Einst}, which is a first particular integral of this system, can be considered as the equation of the \emph{Einstein hypersurface} $\Sigma_E\subset \mathbb{R}_4$ which is, from the other hand, a hypersurface of zero total energy, on which all the phase trajectories of unperturbed cosmological model lie (see \cite{Ignat20}). This equation determines the initial value of the Hubble constant given the other dynamic variables $\Phi,Z=\dot{\Phi}, a$. Let us note that the specified system of equations is easier to study ont the timescale of physical time
\begin{equation}\label{t}
t=\int a(\eta)d\eta,
\end{equation}
since in this case the explicit dependence on the time variable formally disappears in the field equations and the system of equations becomes autonomous \cite{YuI_20} and takes the form \cite{YuI_20}:
\begin{eqnarray}\label{Eq_S_t}
\mathrm{e}\biggl(\ddot{\Phi}+3H\dot{\Phi}\biggr)+m^2\Phi-\alpha\Phi^3= -8\pi\sigma_0(\eta);\\
\label{11-44_t}
\dot{H}+\frac{e\dot{\Phi}^2}{2}+\frac{4}{3\pi}m^4_*\psi^3\sqrt{1+\psi^2}=0,\\
\label{Surf_Einst_t}
3H^2-\tilde{\Lambda}-\frac{q^4\Phi^4}{\pi}F_2(\psi)-\nonumber\\
\frac{e\dot{\Phi}^2}{2}-\frac{m^2\Phi^2}{2}+\frac{\alpha\Phi^4}{4}=0;
\end{eqnarray}
where it is now $\dot{f}=df/dt$ and $H=\dot{a}/a$. However, in fact, the explicit dependence from the time variable, more precisely, from the scale factor $a(t)$, is preserved in the field equations by means of the function $\psi(\eta)$ \eqref{psi(eta)}
However, in our article we used the temporary variable $\eta$ for the adequacy of the standard Lifshitz perturbation (see e.g., \cite{Land_Field}). Further, we will use the notation of  $\dot{f}$ for the derivative over time variable $\eta$.

\section{The Equations of the First Order over Perturbations}
\subsection{Perturbations}
Let us write out the metrics with gravitational perturbations in the following form (see e.g., \cite{Land_Field}):
\begin{eqnarray}
\label{metric_pert}
ds^2=ds^2_0-a^2(\eta)h_{\alpha\beta}dx^\alpha dx^\beta;
\end{eqnarray}
and draw attention to the conformal factor $-a^2(\eta)$  in front of the covariant amplitudes of the disturbances, which disappears for mixed
component of the perturbations $h^\alpha_\beta$.  The covariant perturbations of metric are equal to:
\begin{equation}\label{dg}
\delta g_{\alpha\beta}=-a^2(t)h_{\alpha\beta}.
\end{equation}
Then:
\begin{eqnarray}\label{defh1}
 h^\alpha_\beta=h_{\gamma\beta}g^{\alpha\gamma}_0\equiv-\frac{1}{a^2}h_{\alpha\beta};\\
 \label{defh2}
h\equiv h^\alpha_\alpha\equiv  g^{\alpha\beta}_0h_{\alpha\beta}=
 -\frac{1}{a^2}(h_{11}+h_{22}+h_{33}).
\end{eqnarray}
Further we consider only longitudinal perturbations of the metric, bearing in mind the problem of gravitational stability of plane perturbations, for definiteness directing the wave vector along the $Oz$ axis. We have in this system of coordinates:
\begin{eqnarray}\label{nz1}
 h_{11}=h_{22} =\frac{1}{3}[\lambda(t)+\frac{1}{3}\mu(t)]\mathrm{e}^{inz};\nonumber\\
\label{nz13}
h=\mu(t)\mathrm{e}^{inz};\; h_{12}=h_{13}= h_{23}=0;\nonumber\\
\label{nz2}
h_{33}=\frac{1}{3}[-2\lambda(t)+\mu(t)]\mathrm{e}^{inz}.
\end{eqnarray}

As can be seen from the previous formulas, the matter in our model is completely defined by two scalar functions -- $\Phi(z,t)$ and $p_f(z,t)$ and the velocity vector $u^i$. Let us expand these functions in a series in terms of the smallness of the perturbations with respect to the corresponding functions on the background of the Friedmann metric \eqref{ds_0}:\footnote{In order not to overload the formulas with unnecessary indices, we will accept the following renaming $S_0=S(t)$ for the unperturbed values $S_0(t)$.}
\begin{eqnarray}\label{dF-drho-du}
\Phi(z,\eta)\rightarrow\Phi(\eta)+\varphi(\eta)\mathrm{e}^{inz};&\nonumber\\
\varrho(z,t)\rightarrow\varrho(\eta)+\delta(\eta)\mathrm{e}^{inz};&\\
\sigma(z,\eta)\rightarrow \sigma(\eta)+s(\eta)\mathrm{e}^{inz});&\nonumber\\
u^i=\frac{1}{a}\delta^i_4+\delta^i_3 v(\eta)\mathrm{e}^{inz},&\nonumber
\end{eqnarray}
where $\varphi(\eta),\delta(\eta),s(\eta),v(\eta)$ are functions of the first order of smallness in comparison with their unperturbed values.

\subsection{The Equation for the Perturbations of the Scalar Field}
Expanding the field equation in a Taylor series, we obtain equations for the first-order scalar field perturbations $\varphi$:
\begin{eqnarray}\label{Eq_dphi}
\ddot{\varphi}+2\frac{\dot{a}}{a}\dot{\varphi}+\bigl[n^2+a^2(m^2-3\alpha\Phi^2)\bigr]\varphi\nonumber\\
+\frac{1}{2}\dot{\Phi}\dot{\mu}=-8\pi a^2s.
\end{eqnarray}
This equation differs from the similar one for perturbations of the vacuum scalar field only by the term with the source of the scalar field on the right-hand side (see \cite{YuI_TMP}).

\subsection{The Equations for the Gravitational Perturbations}
Expanding now the Einstein equations in a Taylor series, we obtain the following independent equations for gravitational perturbations of the first-order $\mu,\lambda$:
\begin{eqnarray}
\label{34}
v=\frac{in}{8\pi a^3(\varepsilon+p)_p}\biggl(e\varphi\dot{\Phi}+\frac{1}{3}(\dot{\lambda}+\dot{\mu})\biggr);\\
\label{44}
8\pi a^2\delta\varepsilon_p=\frac{\dot{a}}{a}\dot{\mu}-\dot{\Phi}\dot{\varphi}+\frac{n^2}{3}(\lambda+\mu)\nonumber\\
-a^2(m^2-\alpha\Phi^2)\Phi\varphi;\\
\label{11-33}
\ddot{\lambda}+2\frac{\dot{a}}{a}\dot{\lambda}-\frac{1}{3}n^2(\lambda+\mu)=0;\\
\label{11+22+33}
\ddot{\mu}+2\frac{\dot{a}}{a}\dot{\mu}+\frac{1}{3}n^2(\lambda+\mu)+3e\dot{\varphi}\dot{\Phi}+\nonumber\\
-3 a^2(\Phi\varphi(m^2-\alpha\Phi^2)-8\pi\delta p_p)=0.
\end{eqnarray}
It can be shown (see e.g, \cite{YuI_TMP}), that the differential - algebraic consequences of the equations \eqref{34} -- \eqref{11+22+33} are the equations for the perturbations of the scalar field \eqref{Eq_dphi} and the equations of motion of the first approximation of degenerate matter. The last ones, obviously, are excessive, since the perturbations of the speed and energy density of matter are directly determined by the equations \eqref{34} and \eqref{44}.
From the remaining equations we  choose 3 independent ones: \eqref{Eq_dphi}, \eqref{11-33} and \eqref{11+22+33} in order to maximize the approximation of the mathematical model to the standard Lifshitz theory. In this case, the equation \eqref{11-33} coincides with the corresponding equation of the Lifshitz theory, and the equation \eqref{11+22+33} differs from the corresponding equation from the work \cite{YuI_TMP} only by the material term $\delta p_p$. In the capacity of the equation defining the scalar function $\delta(\eta)$ one can choose the equation \eqref{44}. Thus, we obtain a complete closed system of 4 equations in 4 scalar functions $\varphi,\lambda,\mu,\delta$.

\subsection{Relationship Between Perturbations of Macroscopic Scalars of a Degenerate Statistical System with Perturbations of Fields}
Let us now find the above cited explicit connection between the macroscopic scalars of degenerate Fermi - matter and perturbations of the scalar and gravitational fields.
Taking into account the formulas \eqref{3} and \eqref{4}, valid for an ideal fluid (see \cite{Land_Field}), let us write the perturbation of macroscopic scalars in the first approximation over perturbations:
\begin{eqnarray}\label{dn-dpsi}
\delta n= 3 n(t)\delta(t)\mathrm{e}^{inz}; \\
\delta\psi=\psi(\eta)\gamma(\eta)\mathrm{e}^{inz},
\end{eqnarray}
where in accordance with \eqref{psi} it is
\begin{equation}\label{psi0}
\psi(\eta)=\frac{\varrho(\eta)}{q\Phi(\eta)};\;  n(t)=\frac{\varrho^3(\eta)}{\pi^2}
\end{equation}
and the degenerated function of fermions is introduced
\begin{equation}\label{gamma}
\gamma(\eta)=\delta(\eta)-\varphi(\eta).
\end{equation}
Further, taking into account the relations \eqref{dF} we find the expressions which would be useful for further expression of the perturbations of fermions'  macroscopic scalars:

\begin{equation}\label{s}
s =\frac{q^4\Phi^3(\eta)}{2\pi^2}\biggl(3\varphi F_1(\psi) +\gamma\frac{\psi^3}{\sqrt{1+\psi^2}}\biggr);\end{equation}
\begin{equation}\label{de}\delta\varepsilon_p=\frac{q^4\Phi^4\psi}{2\pi^2}\bigl(\varphi F_2(\psi)+2\gamma\psi^2\sqrt{1+\psi^2}\bigl);\end{equation}
\begin{equation}\label{dp}\delta p_p=\frac{q^4\Phi^4}{6\pi^2}\biggl[\varphi F_2(\psi)+\gamma\frac{\psi^3(2+\psi^2)}{\sqrt{1+\psi^2}}\biggl] ;\end{equation}
\begin{equation}\label{d(e-3p)}\delta(\varepsilon-3p)_p=\frac{q^4\Phi^3}{2\pi^2}\biggl(4\varphi F_1(\psi)+\gamma\frac{\Phi\psi^3}{\sqrt{1+\psi^2}}
\biggr).\end{equation}
As can be seen from the above formulas, all perturbations of macroscopic scalars for a degenerate plasma are completely determined by two functions -- $\varphi(\eta)$ and $\gamma(\eta)$.

Using the relation \eqref{de} in the equation \eqref{44}, we can easily find an expression for the perturbation of the Fermi energy $\gamma(\eta)$ through the perturbations of the fields $\varphi,\lambda,\mu$ and their first derivatives:
\begin{eqnarray}\label{gamma=}
\gamma=-\varphi \frac{F_2(\psi)}{\psi^2\sqrt{1+\psi^2}}+\frac{\pi^2}{q^4\Phi^4\psi^3\sqrt{1+\psi^2}}\times\\
\biggl[\dot{\mu}\frac{\dot{a}}{a}-\dot{\varphi}\dot{\Phi}+\frac{n^2}{3}(\lambda+\mu)-a^2\varphi\Phi(m^2-\alpha\Phi^2)\biggr].\nonumber
\end{eqnarray}
Substituting then \eqref{gamma=} into the relations \eqref{dp} and \eqref{s}, we obtain the required closed system of second order ordinary linear differential equations \eqref{Eq_dphi}, \eqref{11-33} and \eqref{11+22+33} with respect to three functions $\varphi,\lambda$ and $\mu$, which completely determine all physical characteristics of the examined system in the first order of the perturbation theory.

Let us note that in particular, we can obtain the conservation law for the number of particles in the first order of the perturbation theory from \eqref{2a} with an account of \eqref{dF-drho-du}
\begin{equation}\label{dn1=0}
\frac{1}{2}\dot{\mu}+3\dot{\delta}+inv=0.
\end{equation}
This law provides an alternate form of the relation between the perturbation of the Fermi energy and perturbations of the metric and velocity.

\section*{The Conclusion}
Thus, in this article we obtained a complete closed mathematical model of the evolution of plane perturbations in a homogeneous isotropic cosmological scalarly charged degenerate Fermi liquid and established its connection with the previously studied model of perturbation of vacuum Higgs fields. The obtained model was reduced to three ordinary second-order linear differential equations with respect to three functions $\varphi,\lambda$ and $\mu$.

The fundamental difficulty of studying the obtained system of ordinary linear differential equations of the second order is due to the impossibility of obtaining an exact analytical analytical background solution of the model, i.e., - functions $a(\eta)$, $\Phi(\eta)$, $\varepsilon(\eta)$ and $p(\eta)$. The specified problem of finding the background solution was solved by numerical - analytical methods in the above cited work \cite{YuI_20}.

In the next part of the article, we will investigate this model for stability in the short-wave approximation and, thus, we will find out the possible influence of the factor of the scalar charge of matter on the formation of the structure of the Universe.

\subsection*{Funding}

 This work was funded by the subsidy allocated to Kazan Federal University for the
 state assignment in the sphere of scientific activities.

\end{document}